\documentclass[conference, usletter]{IEEEtran}
\usepackage[ruled,vlined,linesnumbered]{algorithm2e}  
\usepackage{amsmath} 
\usepackage{float}
\usepackage{cite}

\usepackage{float}
\usepackage{multirow}
\usepackage{amssymb}
\usepackage{amsmath}
\usepackage{mathrsfs}
\usepackage{amsfonts}
\usepackage{graphicx}
\usepackage{color}
\usepackage{xcolor}
\usepackage{subcaption}

\usepackage{amsmath}

\newcommand\releta{\textit{ReLeTA} }

\begin{document}

\title{ReLeTA: Reinforcement Learning for Thermal-Aware Task Allocation on Multicore}
\author{Shi-Gui Yang, Yuan-Yuan Wang, Di Liu, Xu Jiang, Hui Fang, Yu Yang, Mingxiong Zhao} 
\maketitle 
\begin{abstract}
In this paper, we propose \textit{ReLeTA}: Reinforcement Learning based Task Allocation for temperature minimization. We  design a new reward function and use a new state model to facilitate optimization of reinforcement learning algorithm.  
By means of the new reward function and state model, \releta is able to effectively reduce the system peak temperature without compromising the application performance. We implement and evaluate \releta on a real platform in comparison with the state-of-the-art approaches. Experimental results show  \releta can reduce the average peak temperature by 4 $^{\circ}$C and the maximum difference is up to 13 $^{\circ}$C.

\end{abstract}
\section{Introduction}
The new transistor technology allows more cores to be fabricated on a chip to pursue better performance. 
However, the increased number of cores leads to higher power/energy consumption and undissipated temperature. High operating temperature greatly influences the system performance and reliability 
and exacerbates the chip wear-out. Therefore, effective thermal management techniques have recently attracted a lot attention \cite{Pagani2013Machine}. 

Different approaches were proposed to manage the temperature of a system by means of dynamic voltage/frequency/ scaling (DVFS) \cite{Donald2006}  and dynamic power management (DPM) techniques \cite{Chung1999} or/and a thermal-aware task allocation \cite{Rudi2014Optimum, Shulga2016The, Lu2015, Anup2014Reinforcement}. Recently, machine learning (ML) techniques, like supervised learning and reinforcement learning, have demonstrated its superior ability in prediction and classification. Hence, some works strive to  apply machine learning techniques to conduct power/thermal management and task allocation such as\cite{Shulga2016The, Chung1999, Lu2015, Anup2014Reinforcement}. 
 In \cite{Pagani2013Machine}, the authors comprehensively reviewed all works which deploy ML techniques to conduct power, energy, thermal and resource management.   

ML-based approaches usually adopt supervised techniques, such as linear classification, linear regression prediction, etc, \cite{alpaydin2014a} to predict the system status and classify applications into different categories so that the system scheduler can allocate tasks to a proper core according to ML classification. However, these approaches may suffer from inadequate training data and portability issue, i.e., the model trained on a platform model may not be applicable on different platforms. On the other hand, some approaches exploit reinforcement learning (RL)  \cite{Lu2015, Anup2014Reinforcement} to allocate tasks to cores for effectively managing the system performance/temperature. RL features a \textit{trial-and-error} paradigm, so it does not need any training data. In addition, instead of classifying tasks into different categories, RL derives an optimal decision policy using an on-line learning scheme. Thus, it can be easily applied over different platforms. More details about RL are given in Sec \ref{prel:RL}. Therefore, the RL-based approaches have attracted increasing attention in recent years \cite{Pagani2013Machine}. 

Although the existing RL-based approaches such as \cite{Lu2015, Anup2014Reinforcement} have provided some promising results, these approaches still have two shortcomings: 1) \textbf{Ineffective reward function and state model}: Reward function and state model play a critical role in RL. The existing approach like  \cite{Lu2015} uses a simple reward and state setup which leads to a sub-optimal policy for thermal management, whereas the approach \cite{Anup2014Reinforcement} presents a sophisticated reward which takes into account the system performance and temperature at the same time. However, through experiments, we find that their reward function cannot achieve a good balance between the two metrics (See Fig. \ref{fig:exp2_time_tem} in Sec. \ref{sec:releta}); 2) \textbf{Evaluation on simulators}: instead of evaluating on real platforms, \cite{Lu2015} evaluated their approach on Sniper simulator \cite{carlson2014aeohmcm}. The simulation procedure is time-consuming, and its results cannot accurately reflect all system variance, thereby undermining its applicability on real systems.  

To overcome the above-mentioned issues, in this paper, we propose \releta: \underline{Re}inforcement \underline{Le}arning based \underline{T}ask \underline{A}llocation on multicores for thermal minimization. ReleTA works as an allocator in an operating system, assigning incoming applications to a proper core in order to minimize system temperature. In summary, this paper makes the following contributions:
\begin{enumerate}
	\item We propose a new method to model the system state and design the reward function. The new reward function and state model better facilitate optimization of RL algorithm comparing to the state-of-the-art approaches;
	\item Given the new reward function and state model, we present \releta which  deploys the Q-learning algorithm and neural network to achieve an effective task allocation in terms of thermal management; and
	\item  We evaluate the effectiveness and efficiency of \releta on a real platform in comparison with two existing RL-base approaches and \textit{Linux default scheduler}. The results show \releta can reduce the system temperature by 4$^{\circ}$C on average and 13$^{\circ}$C in the best case while maintaining system performance. 
\end{enumerate}

The reminder of this paper is organized as follows. Sec \ref{sec:related} discusses the related work. Sec \ref{sec:prelim} gives the preliminaries about RL and a motivational example. Sec \ref{sec:releta} presents \textit{ReLeTA}. Sec \ref{sec:exp} demonstrates the experimental results and Sec \ref{sec:conc} concludes this paper. 

\section{Related Work}\label{sec:related}

Machine learning (ML) techniques have demonstrated a huge potential in diverse applications, such as image classification, voice recognition, etc. A plenty of works deploys ML to optimize energy consumption, minimize temperature and manage hardware resource \cite{Pagani2013Machine}. 
In \cite{Shulga2016The}, the authors proposed a task mapping algorithm using Support Vector Machine (SVM) classification \cite{alpaydin2014a} to predict the best task mapping.
In \cite{Chung1999}, the authors used a learning tree to accurately predict the duration of future idle periods to adjust the system's work state.
In \cite{Yang2015Adaptive}, the authors proposed  a task mapping method based on Linear Regression model \cite{alpaydin2014a}.
All these approaches use supervised learning and need a lot of diverse training data to generate an accurate prediction model. However, the diverse training data is difficult to obtain \cite{Pagani2013Machine}. Moreover,  one model trained based on one data set and one hardware model may be inapplicable to other platforms and different applications. A comprehensive review of machine learning on multicore systems can be found in \cite{Pagani2013Machine}

Comparing to the approaches based on supervised learning, RL techniques do not need training data to generate a prediction model. Instead it generates an optimal policy via continuous interactions. Thus, RL-based approaches recently have attracted the increasing attention. Most of them use RL to control the system frequency and operational state such as \cite{Iranfar2015A},  and more can be found in \cite{Pagani2013Machine}. Due to the increasing diversity of applications and underlying hardware platforms, the task allocation on multicore systems has become a challenging problem. Few works stive to adopt RL to allocate tasks for thermal management. 
\releta is very similar to \cite{Lu2015} and \cite{Anup2014Reinforcement} which both optimize the system temperature by RL-based task allocation.
However, the reward function and state model of \cite{Lu2015} are too simple to accurately capture the system execution variation, thereby leading to a sub-optimal task allocator.  In addition, since they evaluated their approach on a system simulator, it is difficult to gain the precise temperature variation during application executions, thus limiting its applicability. Throughout this paper we refer to their approach as \textsf{LTB}.
In \cite{Anup2014Reinforcement}, the authors defined a  multi-objective reward function to guide the system optimization and their RL agent performs the task allocation and frequency scaling at the same time.  However, by our experiments on a real platform, we found their reward and action model cannot derive a good allocation-selection policy with two goals. More details are given in Sec. \ref{prem:motiv}. Throughout this paper, we refer to this approach as \textsf{DSM}. In this paper, we propose \releta to overcome the flaws mentioned above and implement it on a real system.



\section{PRELIMINARIES}
\label{sec:prelim}
\subsection{Reinforcement Learning}
\label{prel:RL}

 \begin{figure}[h]
\centering
\includegraphics[width=0.75\columnwidth]{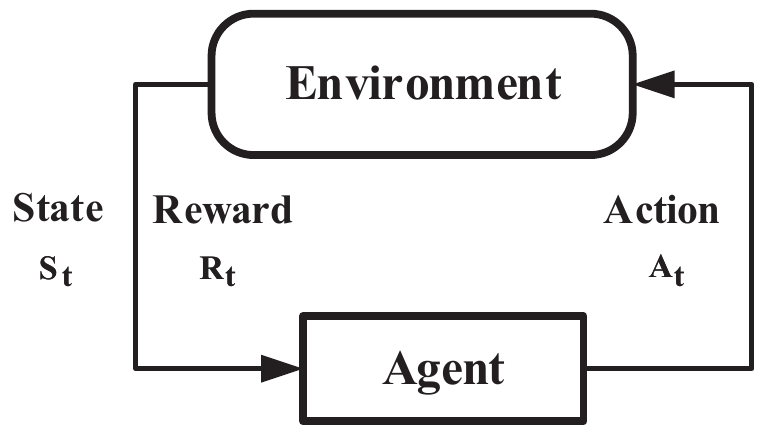}
\caption{Reinforcement learning}
\label{fig:rl}
\end{figure}

 Different from the widely used supervised learning algorithms which learn a prediction model from a plethora of data, RL deploys the \textit{trail-and-error} scheme to `teach' an agent to select an optimal decision in an environment for the maximum reward  \cite{Sutton1998Reinforcement}. RL features an online learning procedure and obtains the optimal decision policy through continuous interaction with the environment.

 The operational semantics of RL is given in Fig \ref{fig:rl}. RL has two components, an \textit{agent} and an \textit{environment}. The agent performs an action based on its current state and the environment responds to the action by returning a reward and states back to the agent. The agent improves its decision policy according to the reward it received. The primary objective of RL is to maximize the long-term accumulated reward.

\subsection{Q-Learning Algorithm}
\label{prel:q-learn}

In this paper, we deploy \textit{Q-learning} algorithm \cite{Sutton1998Reinforcement} as our RL algorithm. \textit{Q-learning} is a promising technique which has been widely used in different contexts. \textit{Q-learning} uses Q value, which is a function of current state $S_i$ and action $A_i$ shown in Eq. (\ref{q_value_update}), to improve its decision policy. The Q value indicates the expected future reward after performing action $A_i$ under state $S_i$ and is updated using the following equation:

\begin{equation}\label{q_value_update}
\small
Q(S_i,A_i)\leftarrow   Q(S_i,A_i)+\alpha \big[R_{i+1}+\lambda \max_{a'} Q(S_{i+1},a')-
Q(S_i,A_i)\big]
\end{equation}
where $\alpha$ and $\gamma$ denote the learning rate and the  discount factor \cite{Sutton1998Reinforcement}, respectively.

%
%
%
%
%

\subsection{Motivational Example}
\label{prem:motiv}
Before proceeding to the presentation of \textit{ReLeTA}, we motivate the proposal of \textit{ReLeTA}. In \cite{Lu2015}, 
they capture all cores' temperature as their system state and calculate the reward using the following equation:
\begin{equation}
\label{eq:lu_reward}
R = T_{em} - T_{\max}
\end{equation}
where $T_{em}$ and $T_{\max}$ denote an estimated maximum temperature and maximum temperature obtained from all cores, respectively.

\begin{figure*}[h]
	\centering
	\begin{subfigure}{0.45\textwidth}
		\centering
		\includegraphics[width=\columnwidth]{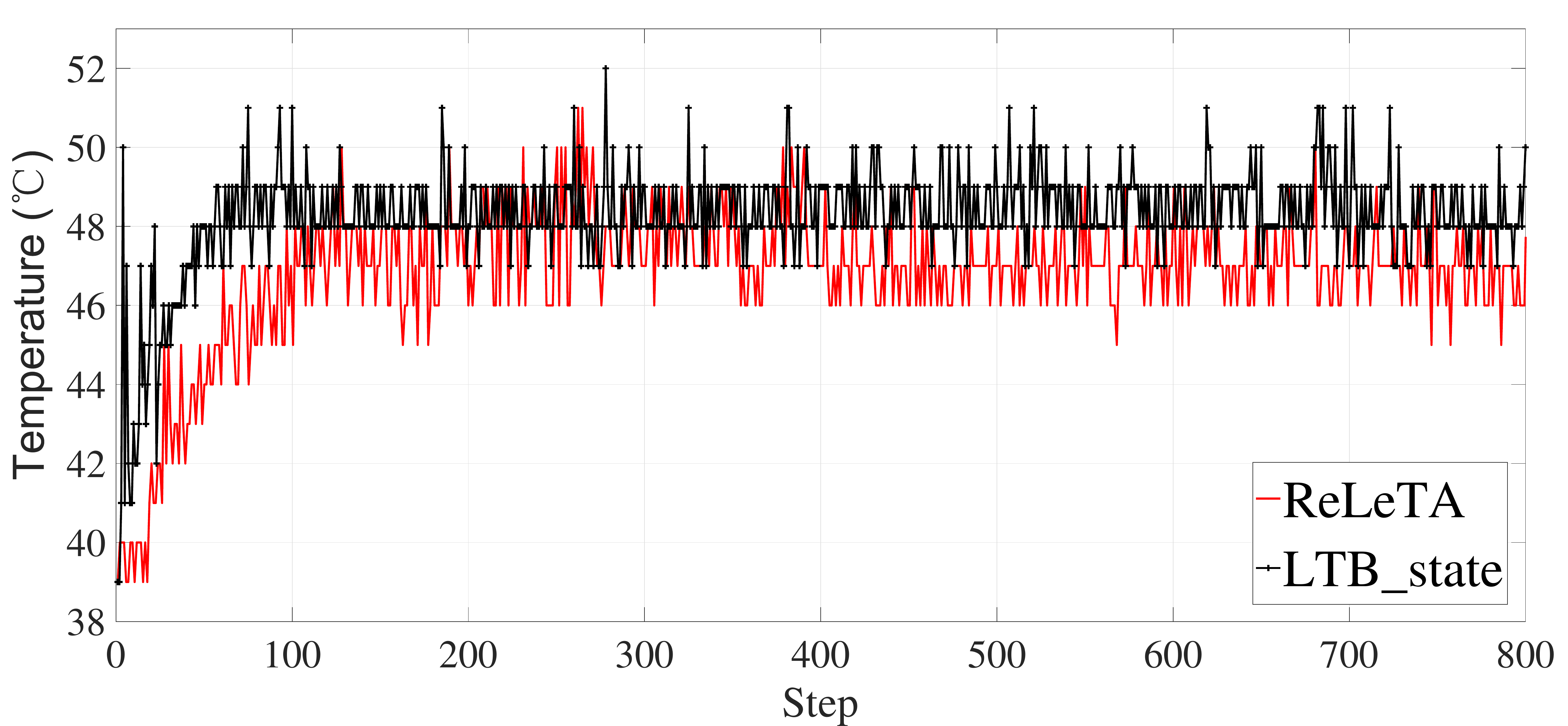}
		\subcaption{Same reward function but different state model}
		\label{fig:compare_state}
	\end{subfigure}
	\begin{subfigure}{0.45\textwidth}
		\centering
		\includegraphics[width=\columnwidth]{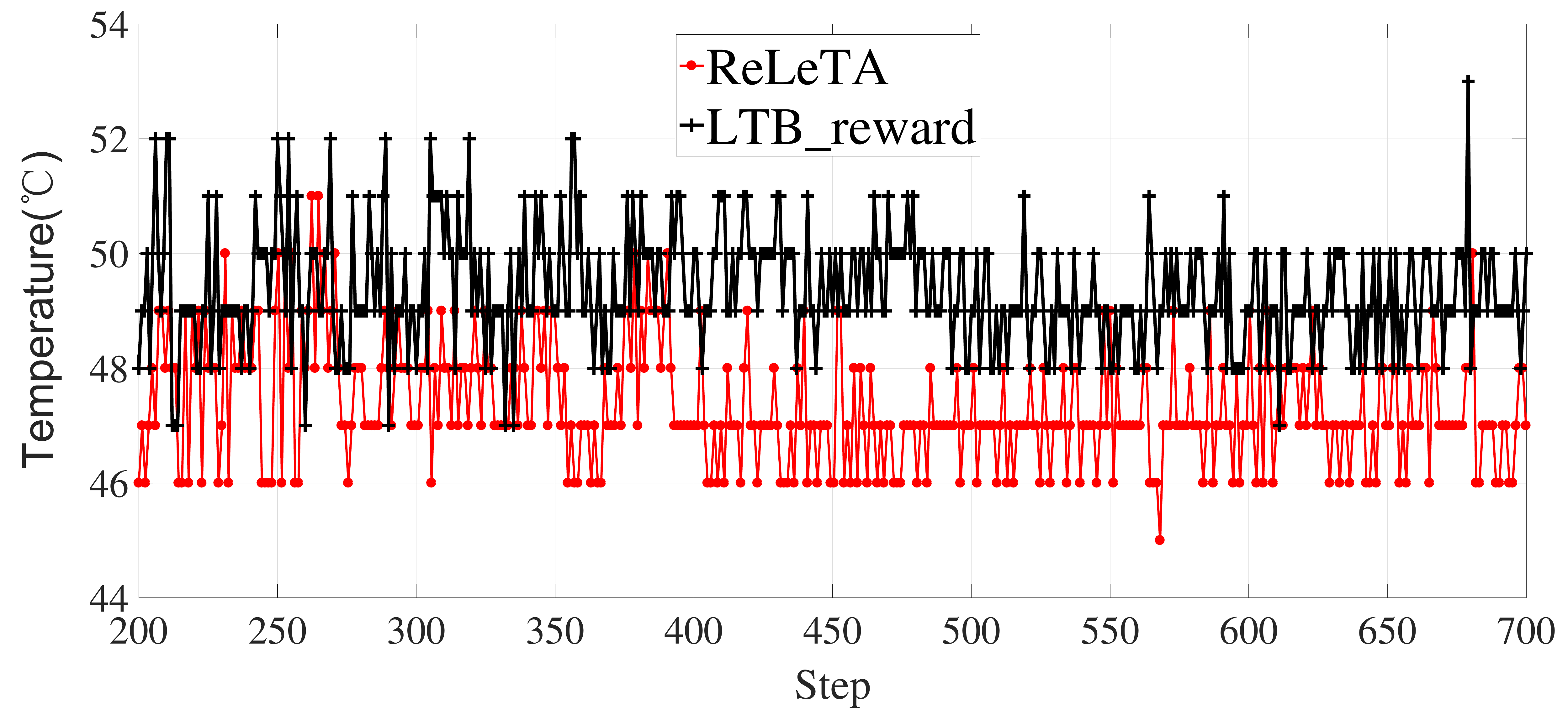}
		\subcaption{Same state model but different reward function}
		\label{fig:compare_reward}
	\end{subfigure}
	\caption{Comparison of different reward and state models}
	\label{fig:motivation}
\end{figure*}
In Fig. \ref{fig:motivation}, we show the experimental results conducted on our experimental platform (See the detail in Sec. \ref{sec:exp})  
in which we use either the same reward function or state model and compare another one with the ones of \cite{Lu2015} in terms of temperature minimization. 
This experiment adopts benchmark \textsf{facesim} from PARSEC benchmark suite \cite{parsec}. We repeatedly execute the benchmarks for 1200 times and record the system temperature variation. Here, we only show the partial results to clearly demonstrate the difference between two approaches. 
When using different state models with the same \releta reward function shown in Fig. \ref{fig:compare_state}, the one with \textsf{LTB}'s state model increases  the system peak temperature by 1.3 $^{\circ}$C on average. 
This is because their state model based on cores' temperature is too simple to accurately capture the system state and workload variation which has a significant effect on  the system temperature.
When using different reward functions with the same \releta state model shown in Fig. \ref{fig:compare_reward}, the one using \textsf{LTB}'s reward function increases the system peak temperature by 2.0 $^{\circ}$C on average.\footnote{Here, we adopt \releta's reward and state, but using \textsf{LTB}'s show the same trend and the difference is slight.} 
We find in some cases different actions (i.e., task allocations)  may have the same $T_{\max}$, so in this situation it is difficult to distinguish the effects of different actions. As a result, this reward function cannot effectively help the algorithm to improve its action-selection policy.

\begin{figure}[ht]
	\centering
	\includegraphics[width=\columnwidth]{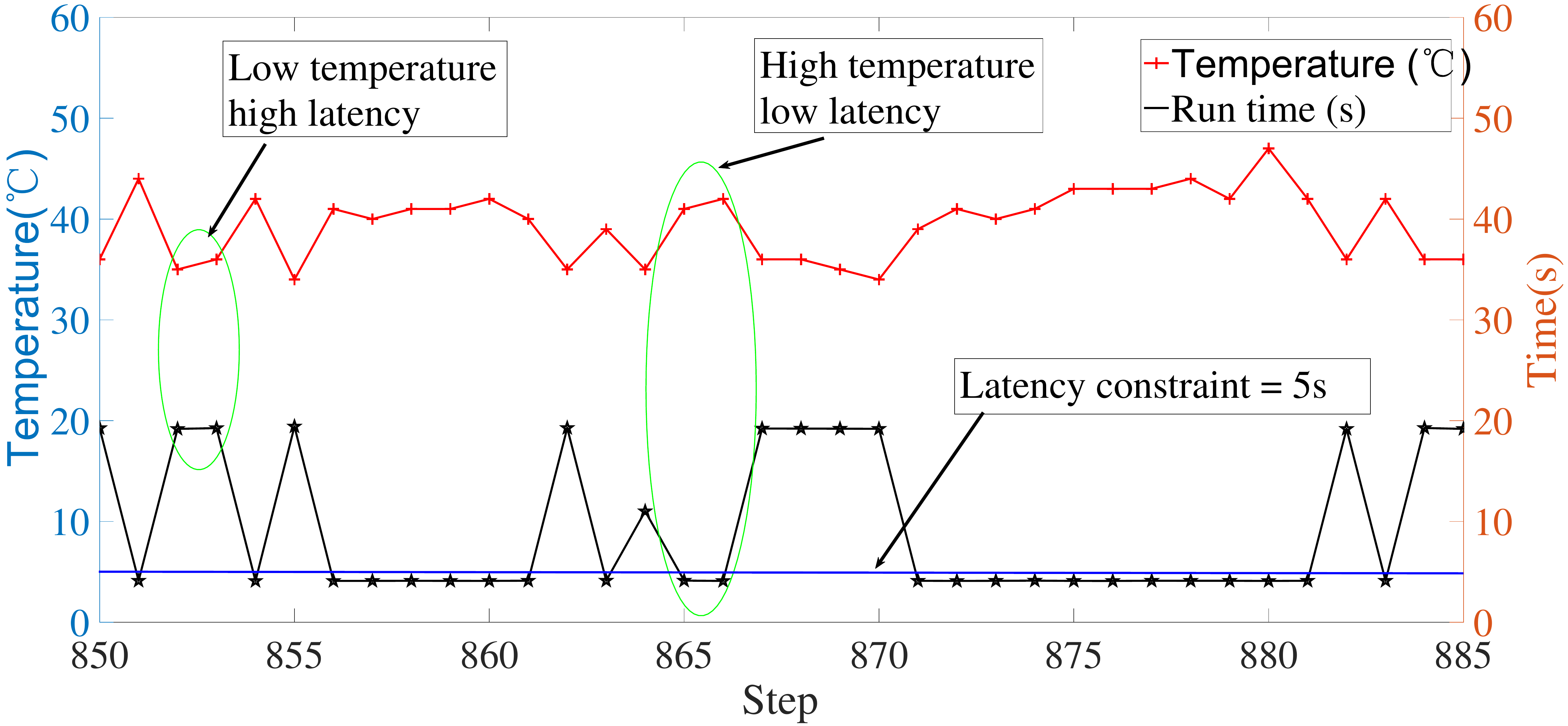}
	\caption{Temperature and time variance of DSM.}
	\label{fig:exp2_time_tem}
\end{figure}

In \cite{Anup2014Reinforcement}, the authors integrated two metrics into the reward function and conducted task allocation and DVFS at the same time. However, we find although their approach can achieve a lower temperature on average, it does not achieve a good balance between two metrics considered. As the consequence, their action-selection policy cannot converge to an optimal policy, and the experimental results are shown in Fig. \ref{fig:exp2_time_tem}. We execute benchmark \textsf{facesim} 1200 times on our experimental platform using \textsf{DSM} approach. Since the reward function of \textsf{DSM} needs a latency constraint,  we specify 5s as its latency constraint. 

Fig. \ref{fig:exp2_time_tem} shows the time and temperature variance spanning from the 850th execution to the 885th execution.  We see that even after more than 800 executions, \textsf{DSM} still misses the specified latency constraint where it may scale down the frequency to purse a lower temperature but completely ignoring the latency constraint, verse vice.  From the examples, \textit{we can see the existing RL-based approaches do not lead to an optimal allocation decision in terms of temperature minimization. Therefore, we present the new \releta.}

\section{ReLeTA}
\label{sec:releta}
\begin{figure}[htbp]
\centering
\includegraphics[width=0.80\columnwidth]{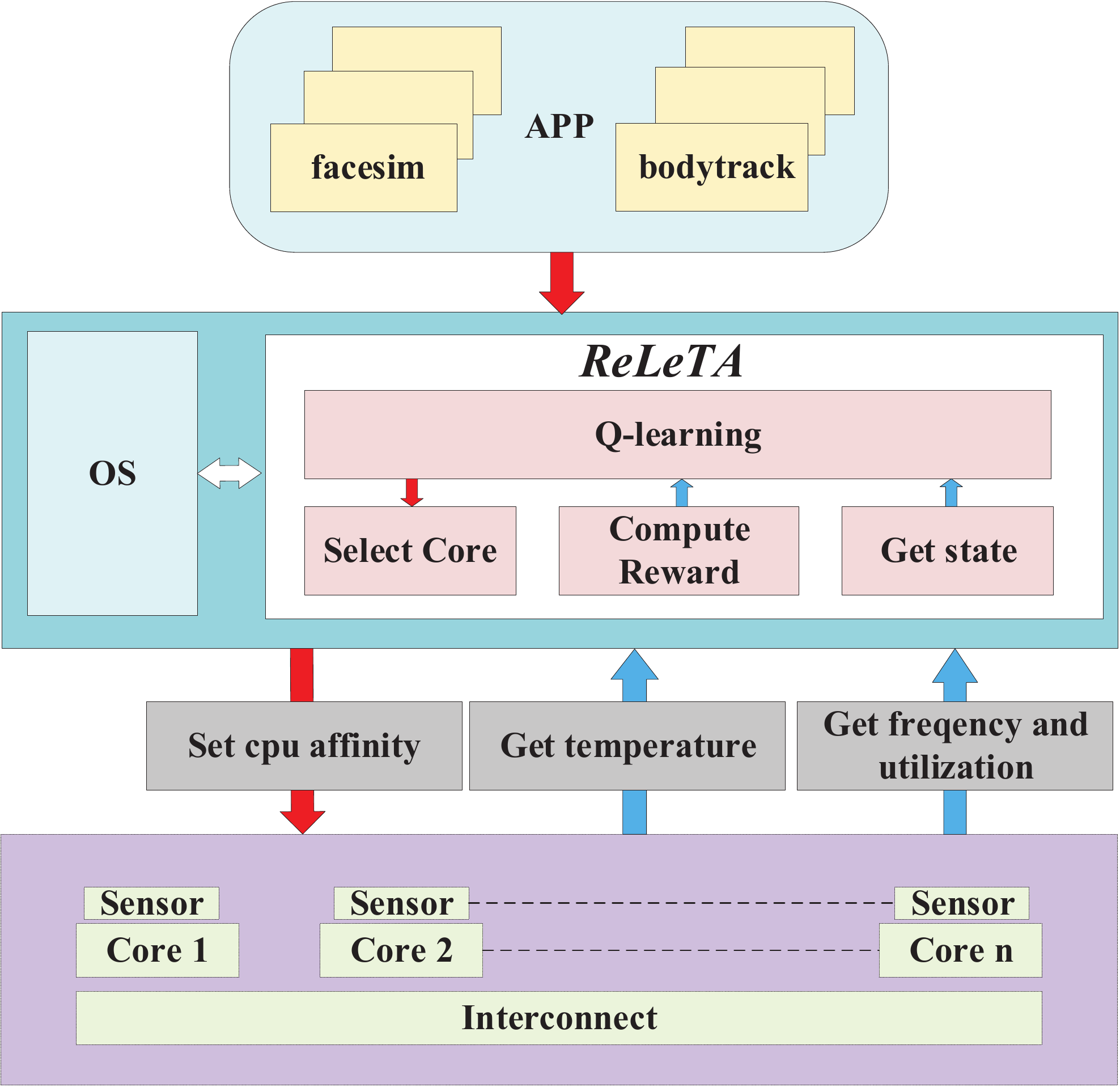}
\caption{The overview of \releta}
\label{fig:system_level_approach}
\end{figure}

Fig. \ref{fig:system_level_approach} presents the overview of \releta, where \releta serves as a task allocater in a system, interacting with OS and obtaining the system information directly from the underlying hardware. When a new application arrives, \releta collects the system state, such as utilization, frequency and temperature from performance counters and system sensors and decides where to allocate the application by using \textit{cpuaffinity} to minimize system temperature. 

As stated in Sec \ref{prel:q-learn}, we adopt \textit{Q-learning} as our fundamental RL algorithm. Therefore, in this section, we present how we setup the action space, design the reward function and model the state for \textit{ReLeTA}. This new reward function and state model facilitates optimization of \releta in terms of temperature minimization. In addition, the traditional \textit{Q-learning} uses a table to store values of all state-action pairs such that the algorithm could find the value through looking up the value table. However, 
the tabular approach is prone to the scalability issue, so in \releta, we design a neural network to effectively approximate Q-value (Details in Sec \ref{releta:nn}). 

\subsection{Action Space}
Since a complex action like \cite{Anup2014Reinforcement} (shown in Fig. \ref{fig:exp2_time_tem}) may lead to unstable results, in \textit{ReLeTA}, we only perform task allocation in each action. Therefore, the action space of \releta is as follows:
\[A=\{1, 2, 3, ...n\}\]
where $n$ is the number of cores the system has. The allocator is the agent in RL, while the system itself is the environment. Every time a new task allocation is done, \releta samples corresponding data from the system as the state and computes the reward for the agent. 


\subsection{State Model}

The allocator makes its decision based on the states received, so the state model should precisely capture the factors which are critical to the system's temperature and may help to find a good task allocation.  

To better model the system variance, we use the combination of each core's frequency and utilization to model the environmental state. The rational behind this new state model is because frequency is closely related to the system temperature and utilization indicates the executing workload which is useful to predict the system status when a new task is pending for allocation. 
Thus, when \releta needs to determine which core to assign an application, it reads the current frequency and utilization of all cores from the system, and combines them into a state vector as follows:
\[S_t=\{F_1 , F_2 , ... F_n, U_1 , U_2 , ... U_n\}\]
where $n$ is the number of active cores on the system. $F_i$ and $U_i$ denote the current frequency level and utilization of core $i$, respectively\footnote{In our case, we assume that there exists a frequency governor like Linux ondemand to help the system control frequencies}. 

%
%
%
\subsection{Reward Function}

Since RL improves its policy through continuous interaction and reward returned by the environment, reward is highly critical to RL. From the motivational example, we see that an ineffective reward function results in a sub-optimal policy in terms of temperature minimization. 


In \textit{ReLeTA}, we use a new way to design the reward function shown in Eq. (\ref{reward_eq}). As seen in Fig. \ref{fig:exp2_time_tem}, although a multi-objective reward function takes into account two metrics, it does not necessarily lead to an optimal policy to achieve a good trade-off between two metrics. Therefore, the new reward function only adopts a single metric, i.e.,  the system temperature. However, instead of using a constant value (the estimated temperature $T_{em}$), we use the average system temperature as the reward reference, which can precisely reflect the real system status. Then, we record the new temperature ($actionTem_t$) of the core after it receives the new task and take as the reward the difference between the reference temperature and the new temperature.

\begin{equation} \label{reward_eq}
R(t)= meanTem_{t-1} - actionTem_t 
\end{equation}
The new reward function directly deploys the real system temperature as reference and the temperature from the assigned core, so it can better evaluate the effectiveness of the allocation performed by the policy. 

\subsection{Q-Value Approximation}
\label{releta:nn}
\releta uses \textit{Q-learning} as RL algorithm to improve the action-selection policy, and the traditional \textit{Q-learning} algorithm uses a table to store all \textit{Q-value}. However, the tabular method suffers from the severe scalability issue, and its memory requirement increases exponentially with the  increasing number of states and actions. 
To mitigate this issue, \textit{neural network} (NN) is deployed to replace the Q-table to approximately estimate the \textit{Q-value} under different states \cite{Sutton1998Reinforcement}. 

In \textit{ReLeTA}, we also adopt NN to approximate \textit{Q-value}. To design an effective NN for our \textit{Q-value} approximation, we empirically evaluate the effect of the number of layers and neuron per each layer. The objective here is to construct a simple NN while not losing its accuracy.  Through a plenty of experiments with different NN settings, we find that a simple three-layer  NN, i.e., one input layer, one hidden layer with $2\times n$ neurons and one output layer network structure, can achieve the same accuracy as some more complex NN structures. Therefore, we select this three-layer NN  as our \textit{Q-value} approximator.  Then, we can use the following equation to approximate \textit{Q-value}:
\begin{equation}
\begin{split}\label{q_function}
Q(s,a)= & \theta_1F_1+\theta_2F_2\cdots+ \theta_nF_n \\
            &+ \theta_{n+1}U_1+\theta_{n+1}U_1 \cdots+\theta_{2n}U_{n}+b
\end{split}
\end{equation}
where $\theta =\{\theta_1, \theta_2,\cdots, \theta_{2n}\}$ denotes the parameters of the NN and $b$ denotes the bias \cite{alpaydin2014a}.

For training the NN, we deploy the following loss function and use stochastic gradient descent (SGD)  to update NN's parameters $\theta$.

\begin{equation}\label{q_loss1}
\begin{split}
L(\theta)=E[(r+\gamma maxQ(s~',a~',\theta)-Q(s,a,\theta))^2]
\end{split}
\end{equation} 
This loss function calculates the difference between the estimated \textit{Q-value} and the true \textit{Q-value}. 

\noindent \textbf{Learning rate}:  Learning rate is a hyper-parameter in NN training which influences the convergent efficiency of the algorithm. Through our experiments, we observe that learning rate of 0.8 shows the best result in terms of the convergent efficiency. Therefore, in this work, we adopt 0.8 learning rate to update the NN.

\subsection{$\epsilon$-greedy Strategy}

In \textit{ReLeTA}, we also adopt the widely used $\epsilon$-greedy strategy to prevent the training policy from converging to a sub-optimal policy \cite{Sutton1998Reinforcement}.  The $\epsilon$ is a probability indicating there is a probability $\epsilon$ the agent randomly selects an action instead of using the action decided by the policy.  This strategy ensures that every action has a small probability to be performed such that the obtained policy will not converge to a sub-optimal one. In \textit{ReLeTA}, we deploy a decay $\epsilon$, i.e., the value of $\epsilon$ gradually decreases with more executions. This means at the initial stage the algorithm is encouraged to explore more possibility to find a better policy, while at the later stage the algorithm is preferred to use its trained policy. $\epsilon$ is set to $0.1$ at beginning and finally reduces to $0.03$.  
\subsection{\releta}
The pseudo code of \releta is given in Algorithm \ref{alg:Framwork}. The input of \releta is an application and the result is a task allocation made by \textit{ReLeTA}. When a new application is released, \releta uses its policy or $\epsilon$-greedy strategy (Line 2-5) to determine an allocation and performs this allocation using \textit{cpuaffinity} (Line 6). From Line 7 to 10, \releta updates the states and improves the NN.

\begin{algorithm}[h]  
\scriptsize
\caption{ReLeTA} 
\label{alg:Framwork} 
\KwIn{A new task}
\KwResult{The task allocation}

  
 Get frequency and utilization of all cores as
state $s_t$ and decide the allocation;
  
  \If{With probability $\epsilon$}{$a_t  \leftarrow$ Randomly selet an action from the action space.}
  \Else{$a_t=argmax_a Q(s_t,a;\theta)$}
 
 Set application affinity using $a_t$;
 
 Get frequencies and utilizations of all core as $s_{t+1}$ ;
 
Compute reward $r_t$ using Eq. (\ref{reward_eq})
 
 
 Compute $r_t$ through formula (\ref{reward_eq});
 
Update $\theta$ based on Eq. (\ref{q_loss1}) and SGD;
\end{algorithm}

\section{Experiment}
\label{sec:exp}
We have evaluated \releta against \textsf{LTB} \cite{Lu2015} and \textsf{DSM}\cite{Anup2014Reinforcement} as well as \textit{Linux default}  scheduler on a real computer system with Ubuntu 16.04 LTS. The platform has an Intel Core I7-4790 with 4 cores of maximum frequency 3.6 GHz and 8GB memory. For our evaluations, we select 8 benchmarks from the widely used PARSEC benchmark suite \cite{parsec},  \textsf{bodytrack}, \textsf{blackscholes}, \textsf{canneal}, \textsf{dedup}, \textsf{facesim}, \textsf{ferret}, \textsf{fluidanimate}, \textsf{freqmine}.  

%
%

%
%
%
\begin{figure*}[ht]
	\centering
	\begin{subfigure}{0.3\textwidth}
		\centering
		\includegraphics[width=\columnwidth]{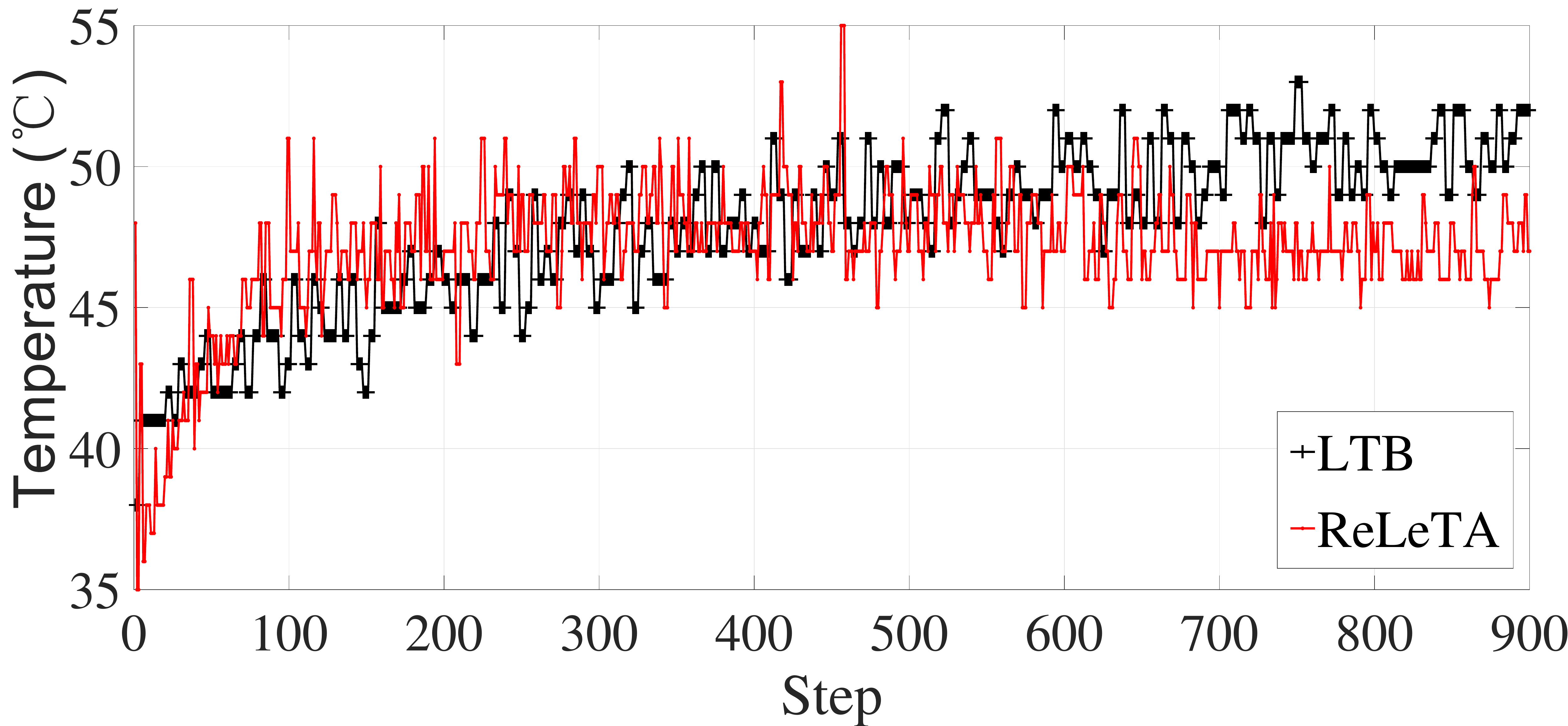}
		\subcaption{3 benchmarks}
		\label{task234}
	\end{subfigure}
	\begin{subfigure}{0.3\textwidth}
		\centering
		\includegraphics[width=\columnwidth]{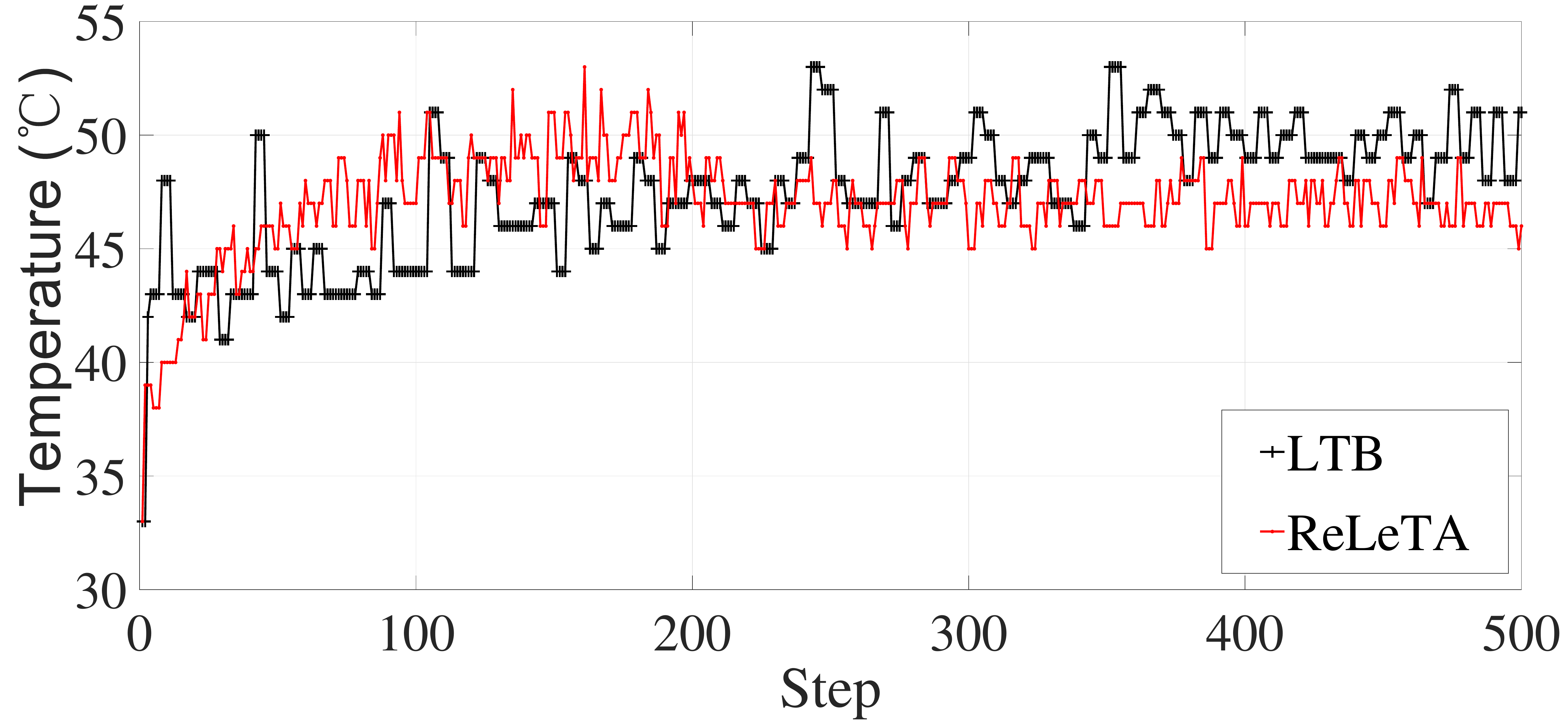}
		\subcaption{5 benchmarks}
		\label{task01234}
	\end{subfigure}
	\begin{subfigure}{0.3\textwidth}
		\centering
		\includegraphics[width=\columnwidth]{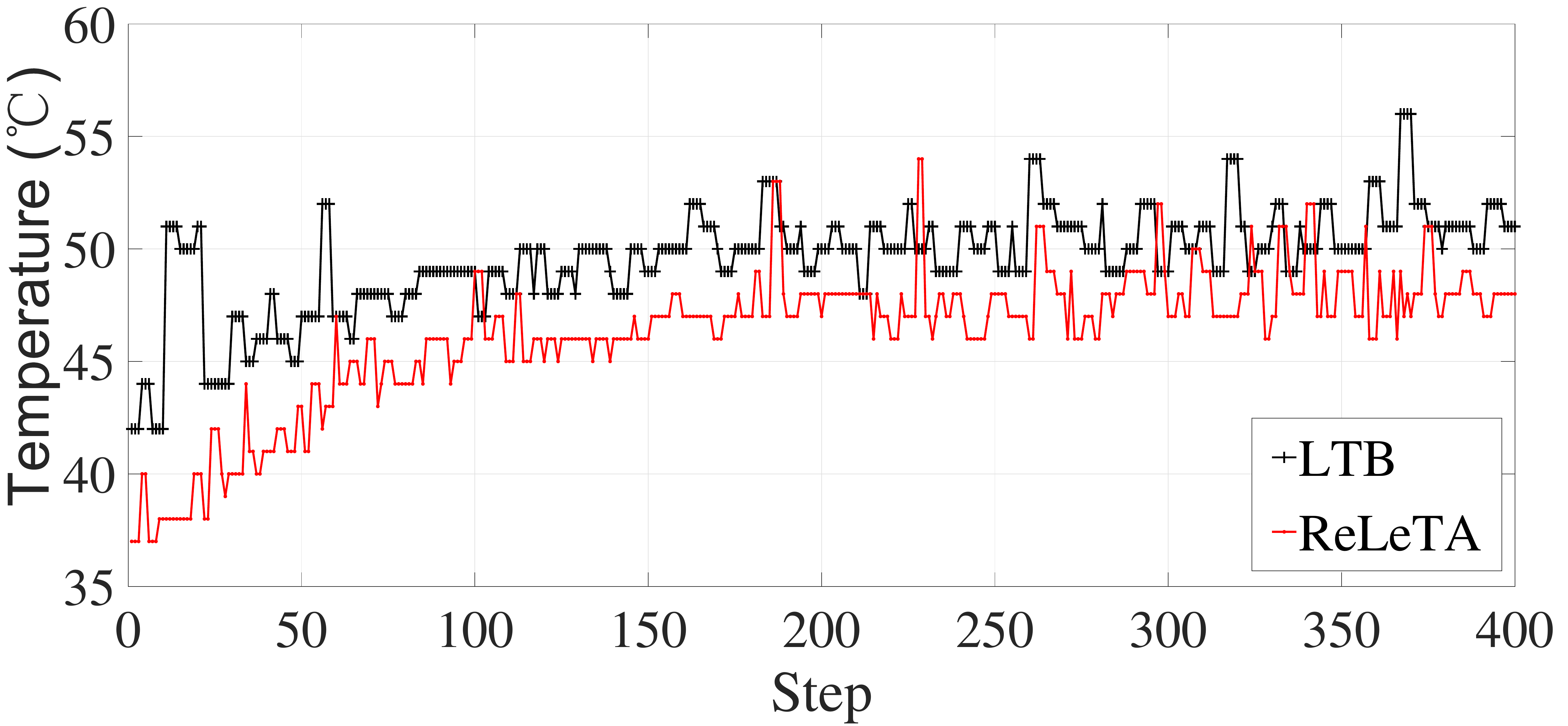}
		\subcaption{8 benchmarks}
		\label{task01234567}
	\end{subfigure}
	\caption{Comparison between LTB and ReLeTA}
	\label{fig:ltb}
\end{figure*}
\subsection{Comparison Between LTB and ReLeTA}
Since \textsf{LTB} has the same objective and execution semantic as \releta, we,  in the first experiment, evaluate \releta  against \textsf{LTB} in terms of temperature minimization. We repeatedly release the benchmarks from the generated  task set with a random time interval. To evaluate the effect of application diversity, we randomly select benchmarks to form three  task sets with 3, 5 and 8 benchmarks, respectively. Both approaches use \textit{Linux governor} `ondemand' to manage CPU frequency.

Fig. \ref{fig:ltb} presents the results using different task sets, where for each experiment, we execute benchmarks in total 2000 times (i.e. each step at \textit{x-axis} implies one application release). The \textit{y-axis} indicates the highest temperature of the system. However, to clearly demonstrate the experimental results, we only show the partial results in which the two approaches have converged to their optimal policies and our \releta has shown its superiority over \textsf{LTB}. From the experimental results, we have the following observations:
\begin{itemize}
	\item When only executing 3 and 5 benchmarks \textsf{LTB} performs slightly better than \releta at the beginning. This is because \textsf{LTB} uses RBF to approximate Q-value, and RBF is known to have faster convergence\footnote{We also evaluate the effectiveness of RBF using our reward and state models, but the NN used in \releta shows better results.}. With the increasing interactions with the system, \releta gradually shows its superiority than \textsf{LTB}. 
	\item When more benchmarks are added to the evaluation, \releta completely outperforms \textit{LTB}. In Fig. \ref{task01234567} \releta demonstrates its superiority over \textit{LTB} in terms of temperature minimization from the beginning.
\end{itemize}
 Table \ref{table:ltb} summaries the numerical results of this experiment, where we see in the best case \releta reduces the average system temperature by 4 $^{\circ}$C when allocating 8 different benchmarks. The maximum difference is 13 $^{\circ}$C when allocating 5 benchmarks.

 \begin{table}
 	\centering
 	\begin{tabular}{|l | l | l|}
 		\hline
 		Task set & Average Diff ($^{\circ}$C) & Max Diff ($^{\circ}$C)\\
 		\hline
 		3 benchmarks & 2.6 $^{\circ}$C & 10 $^{\circ}$C\\
 		\hline 
 		5 benchmarks &  3.7 $^{\circ}$C & 13 $^{\circ}$C\\
 		\hline 
 		8 benchmarks & 4.0 $^{\circ}$C & 8 $^{\circ}$C\\
 		\hline
 	\end{tabular}
 	\caption{Summary of \textsf{LTB} experiments}
 	\label{table:ltb}
 \end{table}

\subsection{Comparison Between DSM, Linux, LTB and ReLeTA}

In the second experiment, we evaluate \releta against \textsf{DSM}, \textsf{LTB} and \textit{Linux} default scheduler. Since \textsf{DSM} performs an action with task allocation and frequency scaling at the same time, other three approaches use \textit{Linux} DVFS governor, `\textit{on-demand}',  to manage each core's frequency.  We use benchmarks \textsf{canneal}, \textsf{dedup}, and \textsf{facesim} to conduct this experiment. In \textsf{DSM} reward function, a latency constraint needs to be specified, so we measure the execution times of three benchmarks and set $2s$, $43$, and $5s$ as their latency constraints, respectively. In total, we execute these benchmarks 400 times. 
\begin{figure*}[htbp]
\centering
\includegraphics[width=0.9\textwidth]{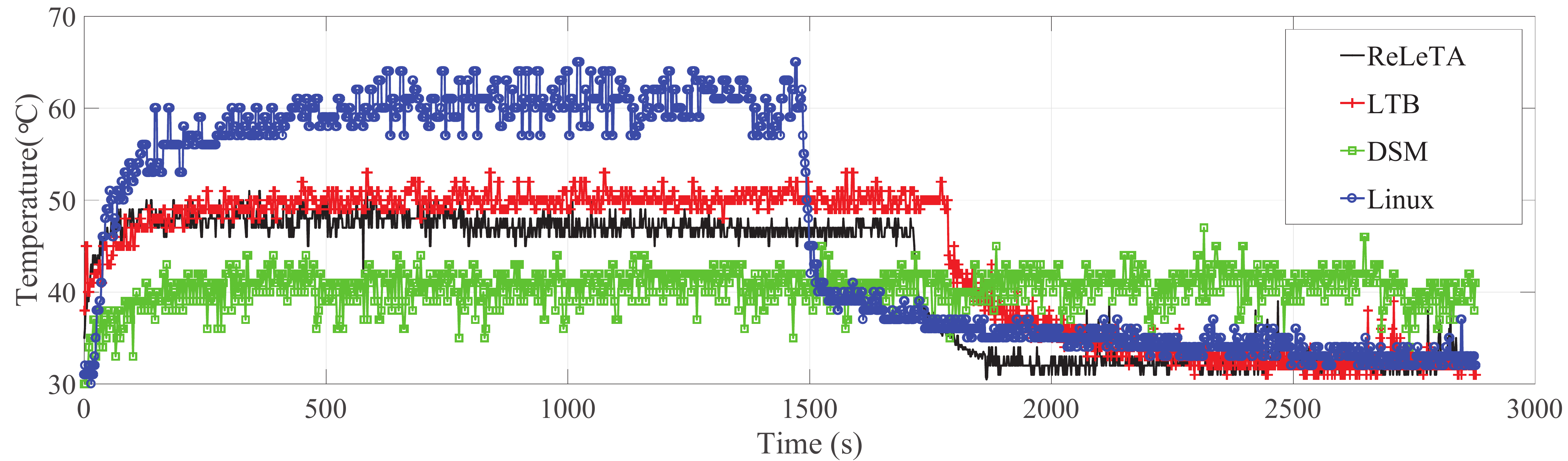}
\caption{Comparison between the four approaches within the same time interval}
\label{fig:all_results}
\end{figure*}

The experimental results are given in Fig. \ref{fig:all_results}, where y-axis shows the peak temperature and x-axis shows execution time. From the experimental results, we find:
\begin{itemize}
	\item Since \textsf{DSM} performs task allocation and DVFS at the same time, it can trade off its execution time for the lower temperature. Hence, we see that \textsf{DSM} completes all execution around $3000s$ with lower temperature, whereas other approaches finish all executions much earlier with higher temperature. Upon the completion, \textsf{LTB}, \textit{Linux} and \releta switch to the power save mode, thereby significantly lowering the system temperature. However, we compare the average system temperature of four approaches over the \textsf{DSM} execution interval, where the results are summarized in Table \ref{tab:temp}. 
	The average temperature of \releta is only 0.6$^{\circ}$C higher than \textsf{DSM}'s. 
	\begin{table}[h]
		\centering
		\begin{tabular}{l |l | l | l | l}
			\hline
			& \textit{Linux} & \releta & \textsf{LTB} & \textsf{DSM} \\
			\hline
			average temp & 46.4$^{\circ}$C &   41.2$^{\circ}$C & 43.7$^{\circ}$C & 40.6$^{\circ}$C\\
			vs \textsf{DSM} & 5.8$^{\circ}$C&   0.6$^{\circ}$C & 3.1$^{\circ}$C & \   \\
			\hline
		\end{tabular}
	\caption{Temperature comparison between the four approaches}
	\label{tab:temp}
	\end{table}
	 \item Although \textsf{DSM} has a latency constraint in its reward function, but their policy does not guarantee to meet it as shown in Fig \ref{fig:exp2_time_tem}. Table \ref{tab:mean_time_of_three_method} shows the rate of the latency constraint violated by all methods. We see that \textsf{DSM} misses latency constraints by 41\%, 5\% and 4\%, respectively, whereas \releta only misses the latency constraints by 2\%, 3\% and 0\% respectively. 
	 \item Comparing to \textit{Linux} default scheduler, \releta violates few latency constraints but reduces the system temperature by 5.2$^{\circ}$C, achieving a good trade-off between the performance and thermal management.
\end{itemize}
\begin{table}[H]
	\centering
	\begin{tabular}{|c|c|c|c|c|}
		\hline
		task & linux & \releta & \textsf{LTB} &\textsf{DSM} \\
		\hline
		\textsf{canneal} & 0\% & 2\% & 0\% &41\% \\
		\hline
		\textsf{dedup} & 0\% & 3\% & 14\% &5\% \\
		\hline
		\textsf{facesim}  & 0\% & 0\% & 0\% &4\% \\
		\hline
	\end{tabular}
	\caption{Latency constraint violation}
\label{tab:mean_time_of_three_method}
\end{table}

\noindent\textbf{Overhead comparison}: Except the temperature and performance, we also evaluate the efficiency of \textit{ReLeTA}, \textsf{LTB} and \textsf{DSM} in terms of allocation overhead. The results are summarized in Table \ref{table:Overhead}, where we can see that the average overhead of all three approaches are smaller than 1ms, but the maximum overhead of \textsf{DSM} is  3 times higher than \releta and \textsf{LTB}.  
\releta has lower overhead than \textsf{DSM}, but higher than \textsf{LTB}. We found the allocation overhead is mainly determined by the complexity of reward function and state model. 

\begin{table}[h]
	\centering
	\begin{tabular}{|c|c|c|}
		\hline
		Method & Average overhead & Max overhead  \\
		\hline
		DSM & 0.776ms & 3.76ms  \\
		\hline
		LTB& 0.296ms& 1.04ms \\
		\hline
		ReReTA& 0.540ms&1.3ms \\
		\hline
	\end{tabular}
	\caption{Overhead Comparison}
\label{table:Overhead}
\end{table}
\section{Conclusion and Future work}
\label{sec:conc}
In the paper, we propose a new task allocation method based on reinforcement learning, namely \releta, to minimize temperature. \releta presents a new reward function and state model to better optimize the task allocation algorithm. 
The experimental results show the effectiveness and efficiency \releta compared to the other two methods. 
In future, we plan to extend \releta to heterogeneous platforms. 

%

\bibliographystyle{IEEEtran}
\bibliography{bibtex}

\end{document}